\documentclass[aip,chaos,reprint]{revtex4-1}
\usepackage[usenames,dvipsnames]{color}
\usepackage{graphicx}
\usepackage{amsmath}

\begin{document}
\title[]{Can the Brownian diffusion coefficient be reconstructed from Lyapunov exponents?}
\author{I. G. Marchenko}
\affiliation{NSC \lq\lq Kharkov Institute of Physics and Technology\rq\rq, Kharkov 61108, Ukraine}

\author{I. I. Marchenko}
\author{D. Ivashchenko}
\affiliation{NTU \lq\lq Kharkov Polytechnic Institute\rq\rq, Kharkov 61102, Ukraine}

\author{\\J. {\L}uczka}
\author{J. Spiechowicz}
\affiliation{Institute of Physics, University of Silesia, 41-500 Chorz{\'o}w, Poland}

%
\begin{abstract}
We consider an ac-driven particle moving in a spatially periodic and symmetric potential. In the zero-temperature limit, for the analyzed parameter set, its 
 dynamics is non-chaotic  and the particle does not manifest diffusive properties. At non-zero temperatures, the asymptotic long-time motion follows normal (Brownian) diffusion. Recent studies have shown that within tailored parameter regimes, the diffusion coefficient is a quasiperiodic function of the external driving amplitude \cite{march-chaos}. Although no general relation between Lyapunov exponents and Brownian diffusion exists, we demonstrate that the quasiperiodic diffusion coefficient at non-zero temperature can be accurately reconstructed from the maximal Lyapunov exponent of the corresponding deterministic system (at vanishing temperature). We propose an approximate formula for this purpose, which shows good agreement with numerical simulations, although some discrepancies are detected in the vicinity of the local maxima of the diffusion coefficient. Finally, we examine the robustness of the correlation between diffusion and the Lyapunov exponent under variations of the system parameters.   
\end{abstract}
\maketitle
\begin{quotation}  
Lyapunov exponents of deterministic dynamical systems are powerful quantifiers used to characterize their properties. In particular, they help to determine whether the system's behavior is chaotic or non-chaotic. The diffusion coefficient is another crucial quantifier that characterizes transport properties. Is there any link between these two radically different classes of quantifiers? In general, no. However, in some cases, one can reveal this link, and both quantifiers can even be strongly correlated. In this work, we consider a particular non-equilibrium system at a non-zero temperature that exhibits diffusive behavior. In the limiting case of zero temperature, the system is in a non-chaotic regime and no diffusion is detected. Remarkably, the maximal Lyapunov exponent behaves qualitatively in the same way as the diffusion coefficient when certain system parameters are varied. This is an astounding correlation because for non-chaotic systems, the Kolmogorov-Sinai entropy is zero, whereas for stochastic ones it is infinite.
\end{quotation}

\section{Introduction}

Particle diffusion is a process in which the long-time mean square displacement of a particle is a linear function of time. There are two approaches to describing this phenomenon: a more fundamental deterministic one and a more phenomenological stochastic one \cite{vulpiani}. In the first case, one can investigate the relationship between properties at the macroscopic level and dynamical behavior at the microscopic scale; for instance, determining whether there are general conditions required for the onset of large-scale diffusive motion. It is widely known that if a deterministic system is chaotic at the microscopic level (characterized by positive Lyapunov exponents or a positive Kolmogorov-Sinai entropy), it exhibits diffusive behavior \cite{gaspard1}. However, there are examples of systems that are microscopically non-chaotic, yet still manifest diffusive properties \cite{cohen1,cohen2,cohen3,klages1}. In the second approach, a thermostat and thermal fluctuations are incorporated according to the fluctuation-dissipation relation, making the system inherently random rather than deterministic from the outset. As a universal and powerful framework, the Langevin equation formalism has been widely utilized, allowing for the inclusion  of both Markovian and non-Markovian dynamics.    

For normal diffusion described by the celebrated Einstein formula for the force-free Brownian particle, the diffusion coefficient $D$ depends on only two parameters: it linearly increases with increasing temperature  and decreases as the friction coefficient grows. For more complex systems, where particles move in potential fields and are driven by external time-dependent forces, $D$ is a much more complicated function of the parameters, and its analytical form is usually unknown. Therefore, numerical simulations are implemented, yielding results that show non-monotonic and often surprising behavior of $D$. In this paper, we consider diffusion of a particle moving in a symmetric, spatially periodic structure and driven by a time-periodic external force.  In one of our previous papers \cite{march-chaos}, a quasi-periodic dependence of $D$ on the amplitude of the external periodic force was found in a certain parameter regime. This quasi-periodicity  can emerge in many subsets of the four-dimensional parameter space of the system. In parallel, we analyze the maximal Lyapunov exponent of its deterministic counterpart and discover that it behaves in a very similar way as the diffusion coefficient. We explain this coincidence in terms of the attractor structure of the system's phase space, which, in the considered parameter regime, is highly regular.   
 
The relationship between Lyapunov exponents and the properties of systems has been  studied in the literature. In Ref. [\onlinecite{posch}], it was shown that some macroscopic transport coefficients, such as those appearing in the Navier-Stokes equations of hydrodynamics, are related to the sum of all Lyapunov exponents. For fluids, there is a dependence between viscosity and the sum of the two maximal Lyapunov exponents \cite{evans}. The existence of a link between relaxation times and Lyapunov exponents has been investigated in gravitating sheet systems \cite{gouda}, as well as in the dynamics of small particles in a random turbulent flow \cite{horvai}. Here, we present one more example and investigate the strong correlation in the behavior of the diffusion coefficient and the maximal Lyapunov exponent when certain system parameters are varied. 

The paper is organized as follows. In Sec. II, we describe the model under study in terms of the Langevin equation and its deterministic counterpart. In Sec. III, we present the dependence of the diffusion coefficient and the maximal Lyapunov exponent on the amplitude of the external time-periodic force. Sec. IV contains an analysis of the deterministic system. In Sec. V, we explain the mechanism underlying the quasi-oscillations of the diffusion coefficient. In Sec. VI, we consider the Lyapunov exponents, their relationship with relaxation times, and propose an approximate equation for the diffusion coefficient formulated in terms of the Lyapunov exponents. In Sec. VII, we discuss various aspects related to the subject of this work. The appendices contain auxiliary material: the scaling of the Langevin equation, details of the simulations, and an approximate theory known as vibrational mechanics. 

\section{Description of the model}

We consider a Brownian particle moving in a spatially periodic structure and driven by a time-periodic force. The corresponding Langevin equation takes the dimensionless form 
\begin{equation}
	\label{La}
	\ddot{x} + \gamma\dot{x} = -\sin{x} + a\sin (\omega t+\phi_0) +  \sqrt{2\gamma Q} \, \xi(t).
\end{equation}
Note that in this scaling the dimensionless mass is $m = 1$. The parameter $\gamma$ is the friction coefficient  and $Q \propto T$ is proportional to 
temperature $T$  of the system. The coupling of the particle to the thermostat is modeled by $\delta$-correlated Gaussian white noise $\xi(t)$ of zero mean and unit intensity, i.e., $\langle \xi(t) \rangle = 0$ and \mbox{$\langle \xi(t)\xi(s) \rangle = \delta(t-s)$}. The starting dimensional equation is presented in the Appendix A, where the corresponding scaling and dimensionless parameters are defined. 

Moreover, we consider the deterministic dynamics of the particle, which corresponds to the zero-temperature limit $Q=0$, namely, 
\begin{equation}
	\label{deter}
	\ddot{x} + \gamma\dot{x} = -\sin{x} + a\sin (\omega t+\phi_0). 
	\end{equation}
It can be converted into a set of three first-order autonomous differential equations. Therefore, there are three Lyapunov exponents $\{\lambda_1, \lambda_2, \lambda_3\}$. One of the exponents is zero, say $\lambda_3=0$. Furthermore, it is a dissipative system with a negative divergence of the vector field defined by this set of equations and the rate of change of the phase-space volume is equal to the friction constant $\gamma$. Hence, the relation 
\begin{equation}
	\label{lap}
	\lambda_1 +  \lambda_2  = -\gamma
 \end{equation}
holds true \cite{golub}. However, this expression alone does not allow for the derivation of analytical expressions for the Lyapunov exponents. 

The complexity of the underlying dynamics---stemming from the nonlinear force $f(x) = -\sin{x}$ with its corresponding spatially periodic potential $U(x)= -\cos{x}$, the time-periodic force $g(t)= a\sin(\omega t+\phi_0)$ with amplitude $a$ and frequency $\omega$, and thermal fluctuations of intensity $2\gamma Q$---does not allow for an analytical approach, which is currently beyond the scope of established mathematical methods. Therefore, numerical simulations have been employed \cite{spiechowicz2015cpc}, and their methodology is described in Appendix B. 

In the deterministic case (\ref{deter}), the system may be non-ergodic  and, consequently, sensitive to the specific choice of initial conditions: the position $x(0)$, the velocity $v(0)$ of the particle, and the initial phase $\phi_0$ of the ac-driving. Thus, all quantities characterizing the system must be averaged over $\{x(0), v(0), \phi_0\}$ using uniform distributions to eliminate this dependence. However, for any non-zero temperature $Q>0$, the system is ergodic, and the initial conditions do not affect its properties in the long-time stationary regime. Consequently, the mean values of observables can be accurately estimated by averaging any single realization over a sufficiently long time interval. 


The model given by Eq. (\ref{La})  has been studied for decades and used for the analysis of various phenomena in both regular and chaotic deterministic regimes \cite{risken,kautz1996,golub}.  Nevertheless, new phenomena remain to be uncovered in this system, carrying the potential for novel applications and  one can still reveal  remarkable outcomes. Recent examples include a non-monotonic temperature dependence of the diffusion coefficient in normal diffusion regimes \cite{spiechowicz2016njp, marchenko2018} and transient, yet extended, time-dependent anomalous diffusion \cite{spiechowicz2016scirep}, to name just a few.

\section{Diffusion and Lyapunov exponents}

The system described by Eq. (\ref{La}) is an example of a non-equilibrium system that can exhibit normal diffusion characterized by the diffusion coefficient 
\begin{equation}
	\label{diffusioncoefficient}
	D =  \lim_{t \to \infty} \frac{\langle \Delta x^2(t) \rangle}{2t}, 
\end{equation}
where 
\begin{equation}
	\label{msd}
	\langle \Delta x^2(t) \rangle = \langle \left[x(t) - \langle x(t) \rangle \right]^2 \rangle = \langle x^2(t) \rangle - \langle x(t) \rangle^2
\end{equation}
is the mean square deviation (variance) of the particle's position $x(t)$. Here, $\langle \cdot \rangle$ denotes the average over thermal noise realizations, as well as over the initial coordinates $x(0)$ and velocities $v(0)=\dot{x}(0)$ of the Brownian particle, and the initial phase $\phi_0$ of the ac-driving.

In Ref. [\onlinecite{march-chaos}], it was found that in the {\it low-friction regime}, the diffusion coefficient $D$ is a damped quasi-periodic function of the amplitude $a$ of the time-periodic force $g(t) = a\sin{(\omega t+\phi_0)}$. The mechanism responsible for this behavior was explained in Ref. [\onlinecite{march-chaos}]. Here, we consider a different regime, namely, the {\it strong-friction sector}. In the deterministic case (\ref{deter}), for the low-friction regime, there are both locked trajectories, in which the motion is confined to a finite number of spatial periods, and running states, where the motion is unbounded in space. In the strong-friction domain, there are only locked trajectories, and in the noisy system (\ref{La}), the mechanism of diffusion is different.  

As an example, we assume the following parameter values: the friction coefficient is $\gamma =3$, the frequency of the time-periodic force is $\omega = 1.59$, the intensity of thermal fluctuations is $Q=0.1$, and the amplitude of the external force $a$ serves as a control parameter. In the deterministic case ($Q=0$), the system is in a {\it non-chaotic regime} and there is no diffusion ($D=0$). 
 
In the noisy case ($Q>0$), the system exhibits normal diffusion with $D > 0$. In panel (a) of Fig. 1,  we show the results of simulations of the Langevin equation (\ref{La}) and the dependence of $D$ on the driving amplitude $a$. Moreover, in  panel (b) of Fig. 1,  we depict the dependence of the maximal Lyapunov 
exponent $\lambda_1$ on $a$ for the corresponding deterministic system (\ref{deter}). One can immediately observe a strong correlation between the dependence of $D$ and $\lambda_1$ on the driving amplitude $a$. This appears to be a surprising coincidence between a characteristic of the stochastic system (\ref{La}) and another characteristic of its deterministic counterpart (\ref{deter}). In general, there is no evident link between these two distinct characteristics of different systems. In the sections below, we explain the behavior of $D$ and $\lambda_1$, and propose an approximate theory for their 
dependence on the driving amplitude   $a$. 

\begin{figure}[ht!]
\centering
\includegraphics[width=0.7\linewidth]{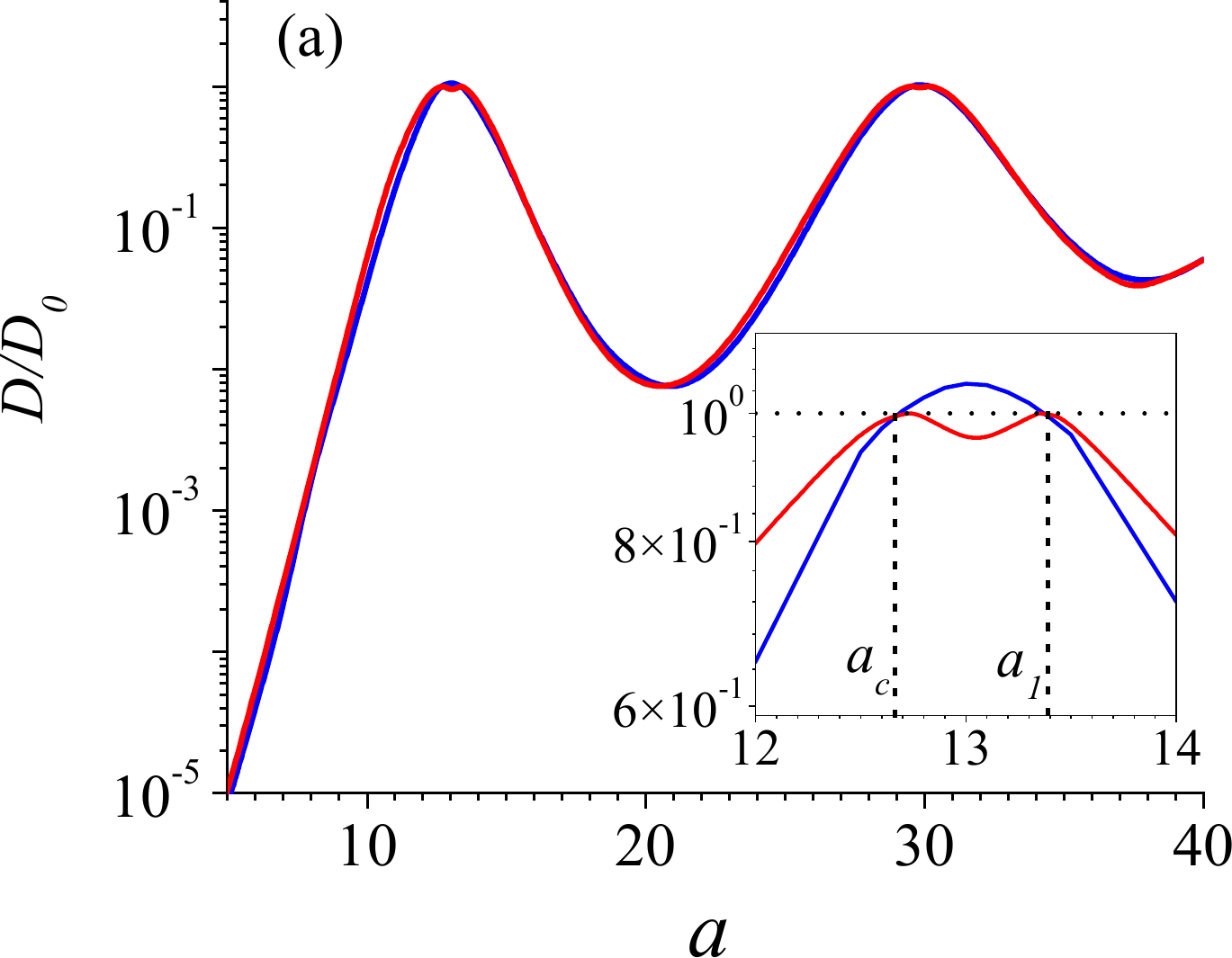}
\includegraphics[width=0.8\linewidth]{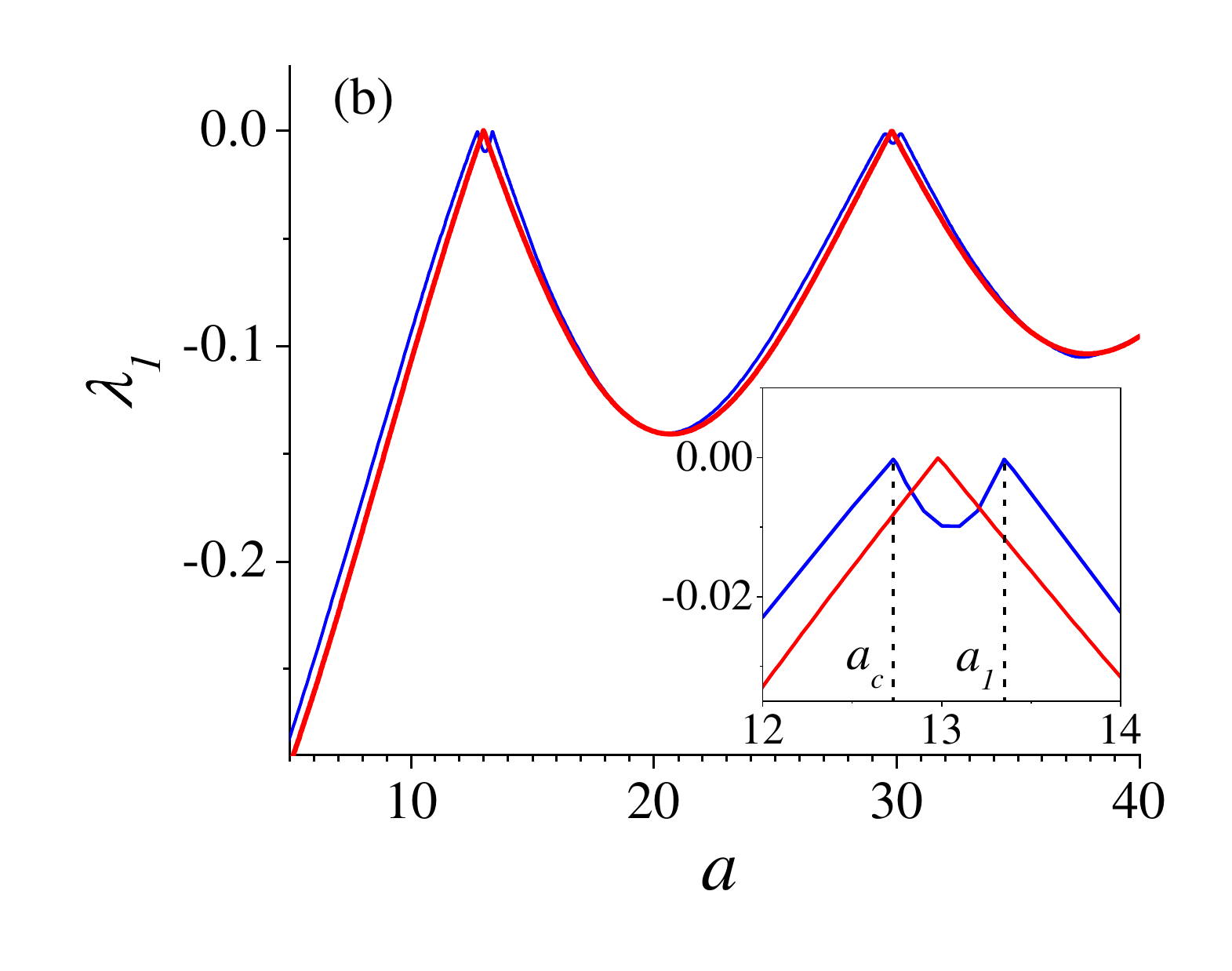}
\caption{(a) The rescaled diffusion coefficient $D/D_0$ as a function of the ac-driving amplitude $a$; here $D_0 = Q/\gamma$ denotes the Einstein diffusion coefficient for a force-free particle. The blue curve represents numerical simulations of Eq. (\ref{La}) and  the red  curve corresponds to the approximate theory (\ref{strongD}). (b) The maximal Lyapunov exponent $\lambda_1$ of the deterministic counterpart ($Q=0$), showing a similar dependence on $a$. 
The blue curve represents numerical analysis of Eq.  (\ref{deter}), while the red curve corresponds to the approximate theory (\ref{root}). Parameters are set to $\gamma = 3$, $\omega = 1.59$, and $Q = 0.1$.} 
\label{fig1}
\end{figure} 

\section{Periodic orbits and bifurcations at $Q=0$}

In the deterministic case and within the strong-damping sector (here, $\gamma = 3$), the stationary states can be characterized by periodic functions of the form 
\begin{equation}
	\label{states}
	x(t) = X_0 + G(t),   
\end{equation}
where $G(t)$ is a periodic function of time and $X_0$ is the center of oscillations. If $x_1(t)$ is a solution of Eq. (\ref{deter}), then $x_2(t)=x_1(t) + 2\pi n$ (where $n$ is an integer) is also a solution of Eq. (\ref{deter}), and the value of $n$ is determined by the initial conditions. However, both solutions will be  identified. i.e. $x_1(t)\equiv x_2(t)$. In  panel (a) of Fig. 2, we present examples of such states. When $X_0 = 0$, the particle oscillates around the minimum of the periodic potential $U(x) = -\cos(x)$. We will refer to this as the normal state, which possesses reflection symmetry. When $X_0 \neq 0$, symmetry-broken states emerge in the system. Below, we follow the analysis presented in Ref. [\onlinecite{japan}]. 

\begin{figure*}[t]
\centering
\includegraphics[width=0.34\linewidth]{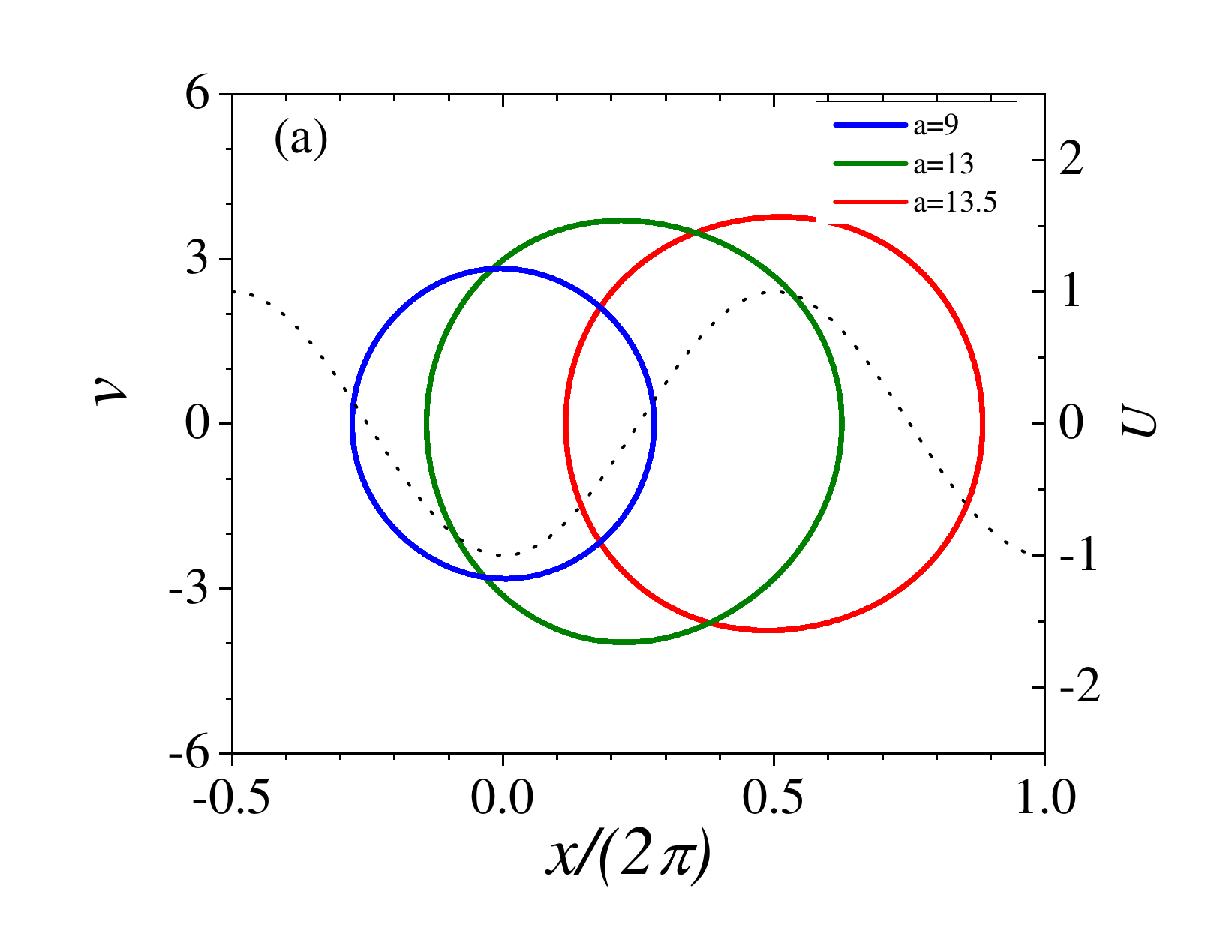}
\includegraphics[width=0.32\linewidth]{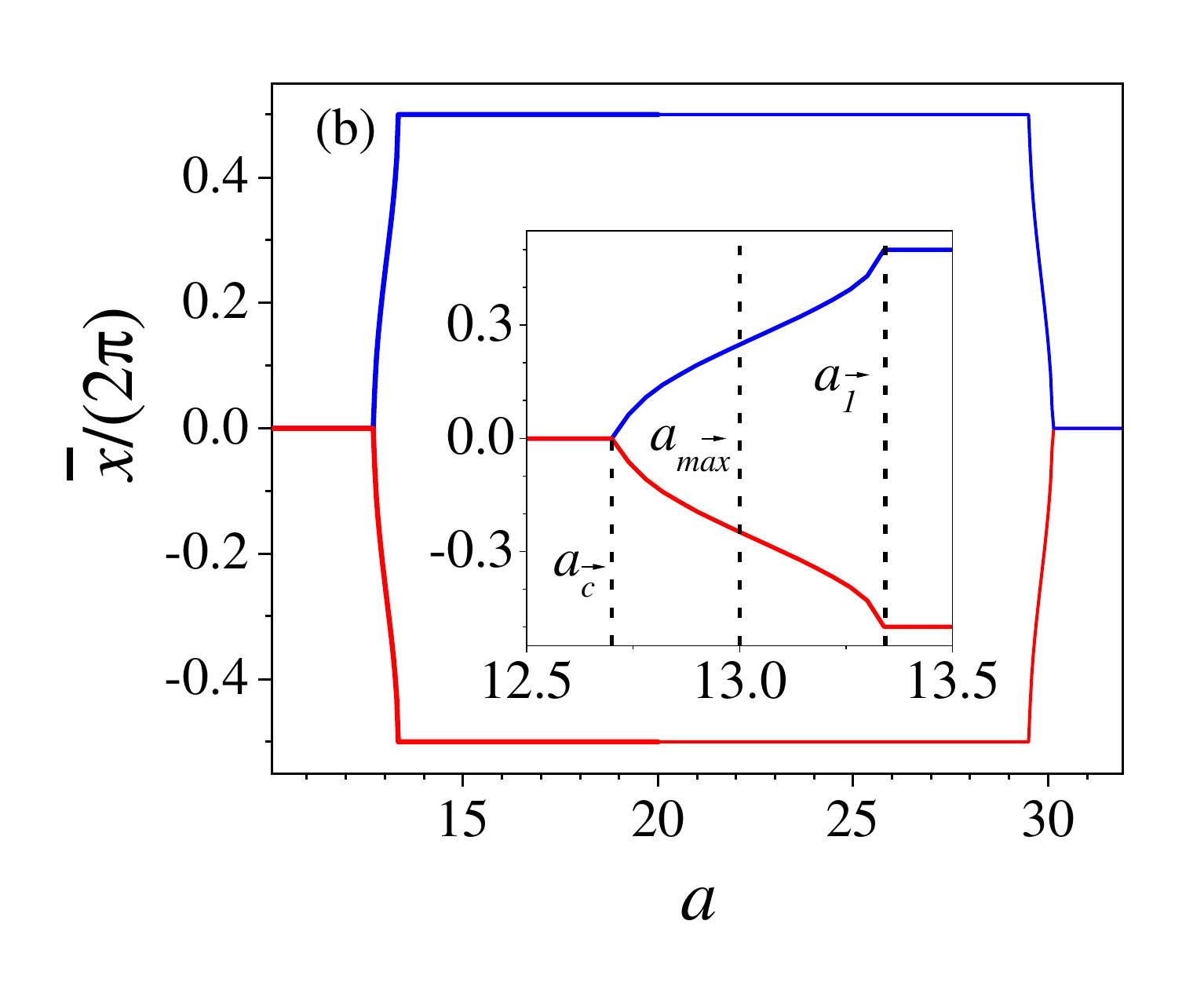}
\includegraphics[width=0.30\linewidth]{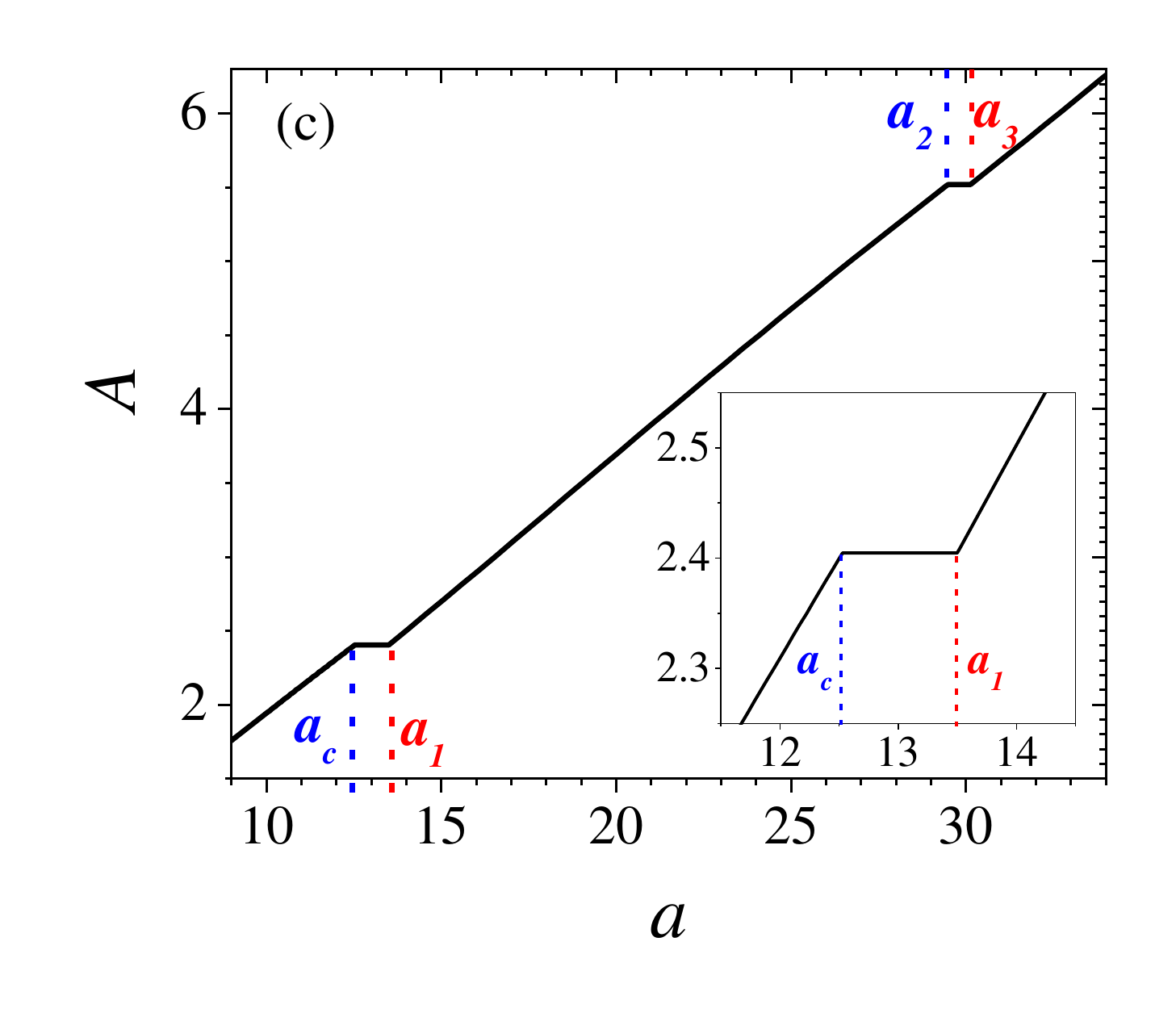}
\caption{Deterministic system ($Q=0$): (a) The phase curves position-velocity in stationary states illustrating transition from normal ($a < a_c$) to symmetry-broken states ($a > a_c$) under variation of the driving amplitude $a$. The dotted curve represents the spatially periodic potential $U(x)$ (b) Bifurcation diagram illustrating the bifurcation point $a = a_c$ where normal and inverted oscillations emerge; $\overline{x}$ is the asymptotic average of $x(t)$ over one period $2\pi/\omega$ of the driving. (c) Amplitude of the periodic orbits versus $a$. Other parameters are the same as in Fig. \ref{fig1}.}
\label{fig2}
\end{figure*}

\subsection{Normal states} 

For a small driving amplitude $a$, the state is normal. In the simplest approximation, the normal state can be represented by the trial function \cite{japan}
\begin{equation}
	\label{normal}
	x(t) =  A \cos(\omega t -\alpha).     
\end{equation}
For a fixed value of $a$, the amplitude $A = A_1$ is determined by the implicit equation
\begin{equation}
	\label{ampli1}
	[2 J_1(A_1) - \omega^2  A_1]^2 +(\gamma \omega A_1)^2 = a^2 
\end{equation}
and $J_1(x)$ is the Bessel function of the first kind \cite{specialF}. The form of the phase $\alpha$ is not important for our analysis  and the influence of higher harmonics is neglected. For example, for $a=9$ and any initial conditions, all trajectories of the particle relax to the attractor, which is a stable periodic orbit depicted in Figs. 2 and 3. In the long-time limit, the particle oscillates around one of the minima of the periodic potential $U(x)$. As the amplitude $a$ increases, the amplitude $A=A_1$ also increases (see Fig. 2, panel (c)). This state remains stable up to the critical amplitude $a=a_c$ determined by the equation 
\begin{equation}
	\label{ac}
	a_c^2 = [2 J_1(A_z) - \omega^2  A_z]^2 +(\gamma \omega  A_z)^2,  
\end{equation}
where $A_z$ is the first zero of the Bessel function $J_0(x)$\cite{specialF}, i.e., $J_0(A_z) = 0$. It follows that $A_z \approx 2.4048$. For the parameters in Fig. 1, i.e., for $\gamma = 3$ and $\omega = 1.59$, the critical amplitude of the external driving is $a_c = 12.53$. Comparing this value with $a_c \approx 12.7$ obtained from simulations, the agreement is reasonably good for the simplest approximation (\ref{normal}).  

\subsection{Symmetry-broken states} 
 
The value $a = a_c$ is a bifurcation point and for $a > a_c$, symmetry-broken states emerge, which are assumed to be approximated by the function \cite{japan}
\begin{equation}
	\label{brok}
	x(t) = X_0 +  A \cos(\omega t -\alpha),    
\end{equation}
where  $A=A_z = 2.4048$  (see Fig. 2, panel (c)) and the center of oscillations $X_0$ is given by the expression   
\begin{equation}
	\label{cent}
	\cos X_0 = \frac{ \omega^2  A_z -  \sqrt{a^2 -(\gamma \omega  A_z)^2} }{2J_1(A_z)}. 
\end{equation}
From this relation, it follows that there are two possible values, $X_0 = \pm |X_0|$, and the sign depends on the initial conditions. If $a = a_c$, then $\cos X_0 = 1$  and consequently $X_0 = 0$. As $a$ increases, $|X_0|$ grows, and the value $\cos X_0 = -1$ is reached at the amplitude $a = a_1$ determined by the equation 
\begin{equation}
	\label{f1}
	a_1^2 = [2 J_1(A_z) +  \omega^2  A_z]^2 +(\gamma \omega  A_z)^2. 
\end{equation}
For the parameters in Fig. 1, it takes the value $a_1 = 13.4998$.

If the amplitude $a$ increases beyond $a_1$, then 
\begin{equation}
	\label{A0pi}
 x(t) = \pm \pi +  A \cos(\omega t -\alpha)
\end{equation}
and the oscillation amplitude $A=A_2$ also increases according to the equation
\begin{equation}
	\label{A1}
	[2 J_1(A_2) +  \omega^2  A_2]^2 +(\gamma \omega  A_2)^2 = a^2.
\end{equation}
This is depicted in Fig. 2, panel (c). It increases up to the amplitude $a = a_2 > a_1$, where 
\begin{equation}
	\label{Azz}
	a_2^2  = [2 J_1(A_{zz}) +  \omega^2  A_{zz}]^2 +(\gamma \omega  A_{zz})^2
\end{equation}
and $A_{zz} \approx 5.52$ is the second zero of the Bessel function $J_0(x)$, i.e., $J_0(A_{zz}) = 0$. For the example in Fig. 1, the amplitude is $a_2 \approx 29.49$. For $a > a_2$, the state is approximated by the function 
\begin{equation}
	\label{Azz}
 x(t) = \tilde X_0  +  A \cos(\omega t -\alpha), 
\end{equation}
where $A=A_{zz}=5.52$ (see Fig. 2, panel (c)) and 
and the oscillation center $|\tilde X_0|$ decreases to zero as 
\begin{equation}
	\label{cent2}
	\cos \tilde X_0 = \frac{ \omega^2  A_{zz} -  \sqrt{a^2 -(\gamma \omega  A_{zz})^2} }{2J_1(A_{zz})}.  
\end{equation}
It follows that $\tilde X_0 = 0$ at the amplitude $a = a_3$ given by the relation 
\begin{equation}
	\label{ampli3}
	a_3^2 = [2 J_1(A_{zz}) - \omega^2  A_{zz}]^2 +(\gamma \omega  A_{zz})^2. 
\end{equation}
In the presented example, $a_3 \approx 30.12$. Consequently, the next normal state is realized again, and a similar scenario is repeated: for $a > a_3$, the oscillation amplitude increases, and the next bifurcation occurs at the third zero of the Bessel function $J_0(A)=0$ for  $A \approx 8.65$, and so on. We summarize this information in  Fig. 2. 

\section{Mechanism of oscillations of diffusion} 

In the previous section, we outlined the approximate theory of the deterministic system (\ref{deter}). In the noisy system (\ref{La}) with $Q > 0$, the picture presented in Sec. IV is blurred out. However, the noisy system `feels' the stable orbits, which play an important role in controlling the diffusion properties. As a simplified picture of the dynamics, one may think of a process in which these orbits are visited in a random sequence. The probability density $p(x)$ for the position of the particle (in the long-time regime, averaged over the initial conditions) takes the characteristic \textit{arcsine} form \cite{levy,arcsine}. Therefore, one still detects a strong concentration of trajectories in the neighborhood of the attractors. The probability of finding the particle residing near the minimal and maximal positions  is particularly large.

In Fig. 1, panel (a), the diffusion coefficient $D$ increases in some intervals of the driving amplitude $a$ and decreases in others. We analyze both cases separately. It is noteworthy that in the interval $[a_c, a_1]$, bounded by two critical values of the driving amplitude $a_c$ and $a_1$, the diffusion coefficient satisfies $D > D_0$, where $D_0 = Q/\gamma$ is the celebrated Einstein diffusion coefficient of a force-free Brownian particle (see the inset in Fig. 1). This interval is a region of  transition between the state (\ref{normal}), characterized by oscillations around the minimum of the potential $U(x)$, and the state (\ref{A0pi}), characterized by oscillations around the maximum of $U(x)$. An example of this is the ``green'' periodic orbit for $a = 13$ in Fig. 2, panel (a).

\subsection{The case $a< a_c$}
 
For a driving amplitude $a < a_c$, the particle oscillates around the minimum of the periodic potential $U(x) = -\cos x$, as shown in Fig. 3, panel (a). Occasionally, the particle is kicked by thermal fluctuations and escapes its initial trajectory, jumping to another trajectory located in a different potential well. After some random time, the particle may again break away from its current trajectory and jump to another one. From Fig. 3, panel (a), it can be seen that for $Q = 0$, the particle moves between two positions within the interval $[-A_1, A_1]$ inside the potential well. The value of $A_1$ increases as $a$ increases, causing the potential barrier to decrease. The minimal potential barrier height can be calculated from the relation $\Delta U(a) = U_{max} - U(A_1)$, where $A_1$ is determined by Eq. (\ref{ampli1}). For the three cases shown in Fig. 3, from the presented theory of the deterministic system (\ref{deter}) it follows that $\Delta U(9) = 0.81$ for $a = 9$, $\Delta U(11) = 0.47$ for $a = 11$, and $\Delta U(12.7) = 0.26$ for $a = 12.7$. More precise values of the barrier heights follow from simulations for  temperature $Q = 0.1$ and take the following values: $\Delta U(9) = 0.82$ for $a = 9$, $\Delta U(11) = 0.48$ for $a = 11$, and $\Delta U(12.7) = 0.24$ for $a = 12.7$. In both approaches, the trend in the variation of the potential barrier is similar. If the barrier  height  is smaller then it is easier for the particle to overcome the potential barrier induced by thermal fluctuations  and jumps to other basins of attraction.  Therefore  $D$ increases as $\Delta U$ decreases. This explains why, in certain regimes, the diffusion coefficient $D$ can increase as the driving amplitude $a$ grows. 

\begin{figure}[t]
\centering
\includegraphics[width=0.8\linewidth]{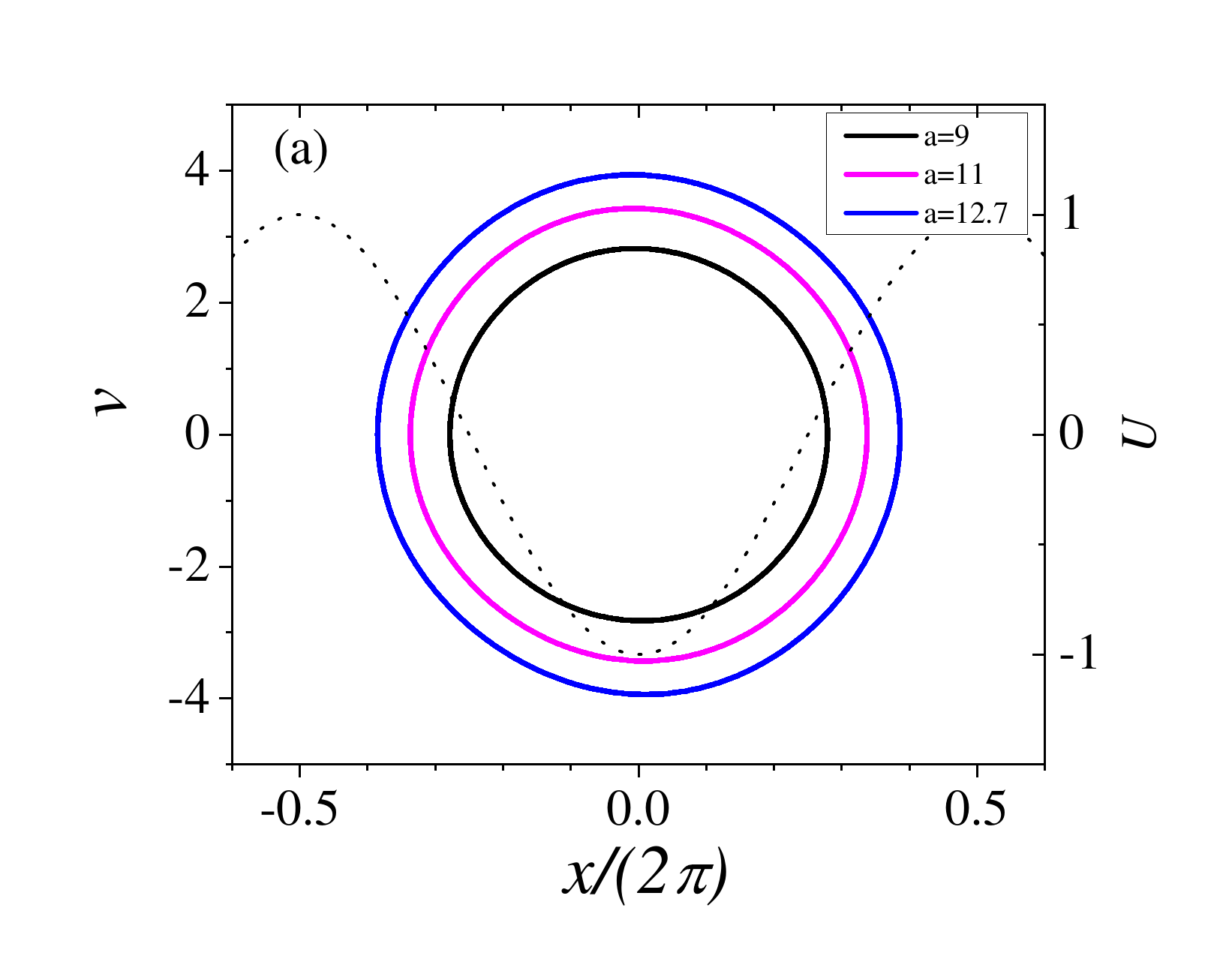}
\includegraphics[width=0.8\linewidth]{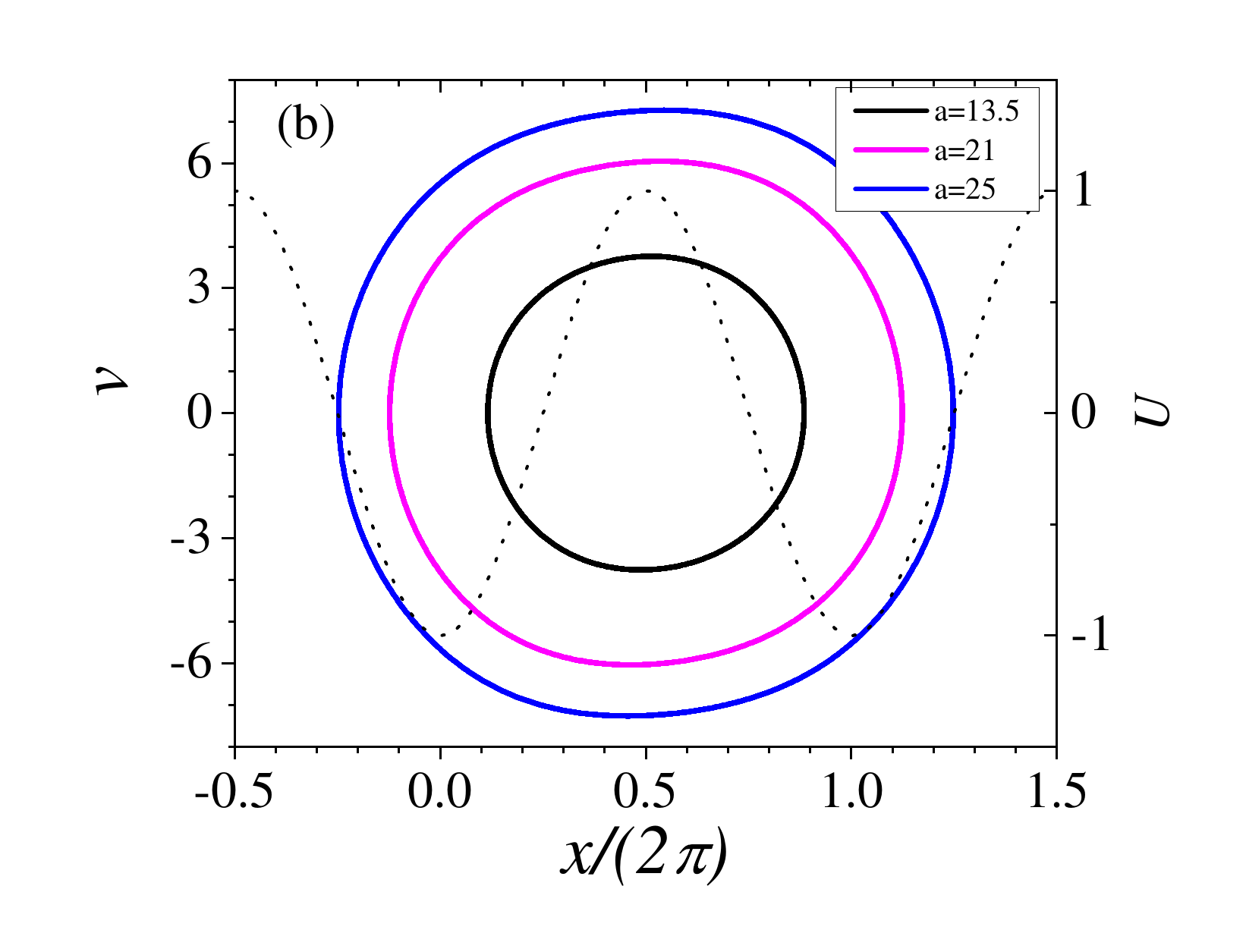}
\caption{Deterministic system ($Q=0$): Periodic orbits for normal and symmetry-broken states illustrating the influence of the external driving amplitude $a$. Remaining parameters are consistent with Fig. \ref{fig1}.}
\label{fig3}
\end{figure} 

\subsection{The case $a> a_c$}

When $a > a_c$, the symmetry-broken states (\ref{brok}) and (\ref{A0pi}) emerge, as shown in  panel (b) of Fig. 3. The oscillation center $X_0$ shifts toward the maxima of the periodic potential $U(x)$, and the dependence of $D$ on $a$ becomes non-monotonic: first, $D$ increases to its maximal value at $a_{max} \approx 13$, then it decreases to its local minimum at $a = 21$. Subsequently, it increases again until reaching the next maximum at $a \approx 30$. For the force amplitude 
 $a = 13$, the particle oscillates in an interval that contains both the minimum and maximum of the potential $U(x)$ (see the green orbit in Fig. 2, panel (a)). When the particle is in the vicinity of the maximum of $U(x)$, it can easily jump to other, more distant wells. This is the primary contribution to the large diffusion coefficient. For the three cases depicted in panel (b) of Fig. 3, it follows that $\Delta U(13.5) = 1.74$ for $a = 13.5$, $\Delta U(21) = 1.73$ for $a = 21$, 
and $\Delta U(25) = 1.03$ for $a = 25$. From simulations of the noisy system (\ref{La}) and for  temperature $Q = 0.1$ one gets the following values: 
$\Delta U(13.5) = 1.75$ for $a = 13.5$, $\Delta U(21) = 1.71$ for $a = 21$, and $\Delta U(25) = 1.02$ for $a = 25$. 
It explains the decrease in $D$ as $a$ grows. For the case $a = 21$, the particle oscillates around the maximum of $U(x)$ but resides mainly near the potential minima (recall the \textit{arcsine} distribution $p(x)$ for the particle position). Consequently, $D$ is minimal. For $a = 25$, the barrier is lower than for $a = 21$, and therefore $D$ is larger. We should be aware of the limitations of our analysis and the presented explanation, which is based primarily on the deterministic picture (\ref{deter}), because the full system (\ref{La}) exhibiting diffusion is noisy and its random dynamics is significantly more complex.

\begin{figure*}[t]
\centering
\includegraphics[width=0.32\linewidth]{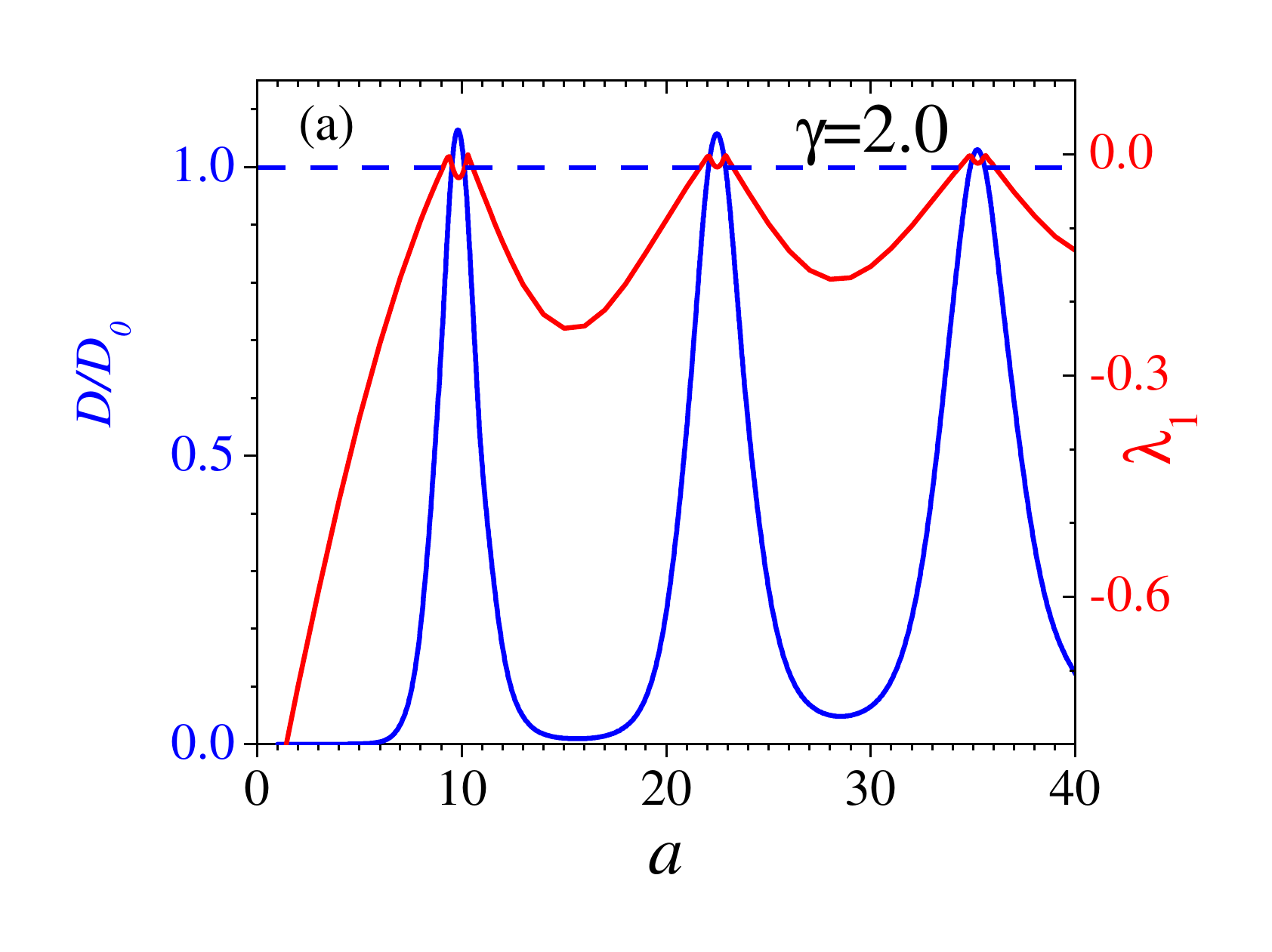}
\includegraphics[width=0.32\linewidth]{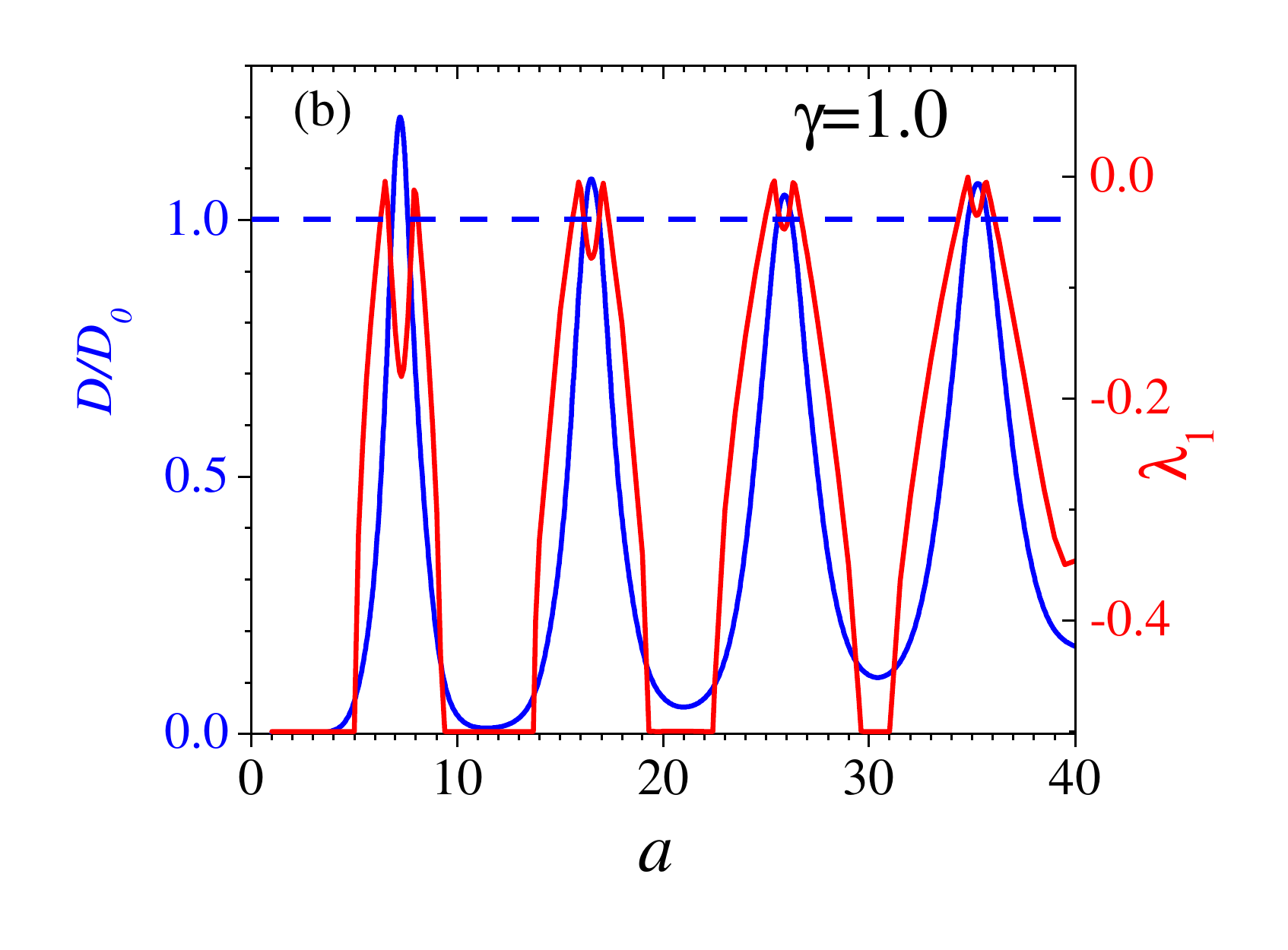}
\includegraphics[width=0.32\linewidth]{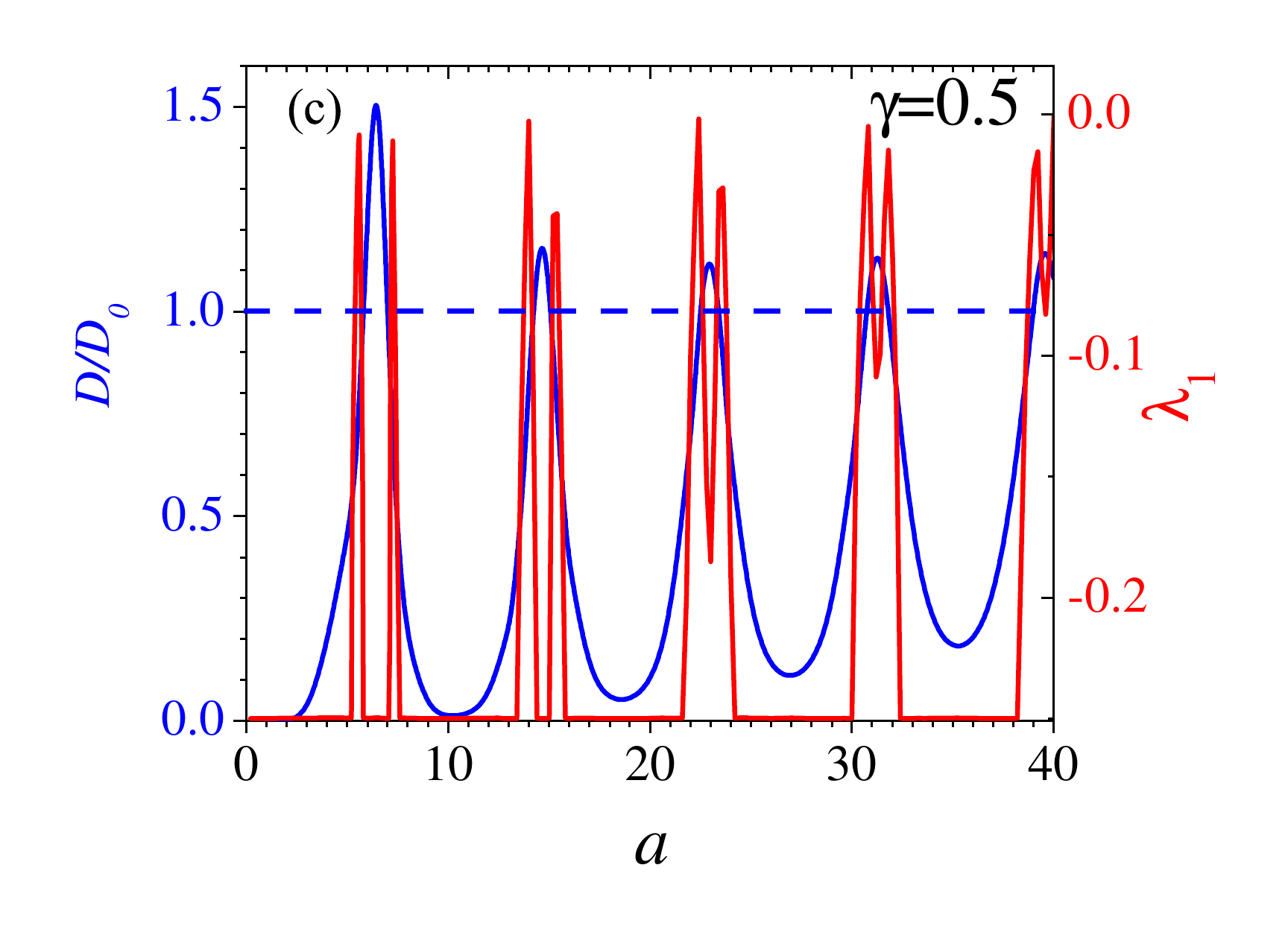}
\caption{Influence of dissipation on the system dynamics for damping constants $\gamma = 2$, $1$, and $0.5$. Other parameters are the same as in Fig. \ref{fig1}.}
\label{fig4}
\end{figure*} 

\section{Mechanism of oscillations of Lyapunov exponent}

The negative Lyapunov exponent $\lambda < 0$ measures the average rate of convergence of neighboring trajectories, and $|1/\lambda|$ is the characteristic time for this convergence. In turn, the relaxation time $\Gamma$ characterizes the approach of trajectories toward a stationary state of the system (i.e., the attractors). Given the strong damping $\gamma = 3$ assumed in this paper, the regime is non-chaotic, and the attractors are periodic orbits. As $t \to \infty$, all trajectories converge to one of them. Therefore, there should be some link between $\lambda$ and $\Gamma$. Below, we elucidate this relationship.

In Fig. 1(b), the blue curve is the result of numerical calculations, while the red curve follows from the theory which we propose here. To this end, we utilize vibrational mechanics (see the Appendix C), which is an approximation of the starting Langevin equation (\ref{La}). Its deterministic version  reads 
\begin{equation}
	\label{VM2}
	\ddot{\tilde x} + \gamma\dot{\tilde x} + k  \sin{\tilde x} = 0, \quad  k = J_0(\psi),  
\end{equation}
where  $\tilde x = \tilde x(t)$ is a slow part of the composite process $x(t)$,  namely, 
\begin{equation}
	\label{sep}
	x(t) = {\tilde x}(t) + \psi \sin(\omega t +\phi_1)
\end{equation}
and  $J_0(\psi)$ is the zero order Bessel function \cite{specialF}. The argument $\psi = \psi(a)$ is defined in Eq. (\ref{psi}). 
In the strong damping sector the solutions $\tilde x(t)$ should relax to zero because in the long-time limit trajectories $x(t)$ approach periodic orbits. 

Eq. (\ref{VM2}) is equivalent to a set of  two autonomous differential equations of first order.  However, analytical formulas for the Lyapunov exponents are still lacking. The only available analytical expression is for their sum \cite{verhulst}, which for Eq. (\ref{VM2}) reduces to the relation (\ref{lap}).  
In order to evaluate the relaxation time of $\tilde x(t) \to 0$,  Eq. (\ref{VM2}) is approximated by the linearized equation ($\sin \tilde x \approx \tilde x$) in the form  
\begin{equation}
	\label{oscilator}
	\ddot{\tilde x} + \gamma\dot{\tilde x} + |k| \tilde x = 0. 
 \end{equation}
We set up  the absolute value of  $k = J_0(\psi)$ in order to ensure a positive ``spring constant'' $k$, yielding an equation for a damped harmonic oscillator in which all trajectories relax to zero. This equation has two characteristic roots 
\begin{equation}
	\label{root}
	\Gamma_{+/-} =  -\frac{1}{2} \gamma \pm \frac{1}{2} \sqrt{\gamma^2- 4 |J_0(\psi)|} .  	
 \end{equation}
The next  assumption is that we equate $\Gamma_{+} = \lambda_1$ and $\Gamma_{-} = \lambda_2$ with two Lyapunov exponents $\lambda_1$ and $\lambda_2$. Then, Eq. (\ref{lap}) is satisfied, and $\Gamma_{+}$ appears to be a good approximation of the maximal Lyapunov exponent $\lambda_1$ shown in Fig. 1. It is remarkable that these premises, which lead to Eq. (\ref{root}), reproduce reasonably well the results of the (exact) numerical analysis.  The key factor in this agreement is the renormalized potential, where the effective potential barrier depends on the amplitude of the time-dependent force and is quantified by $J_0(\psi)$, which behaves like a quasi-periodic decaying function of its argument. However, the approximation (\ref{root}) fails at the critical points. Simulations show that $\lambda_1 \to 0$ as $a \to a_c, a_1, a_2, a_3$. In the proposed theory $\lambda_1 \to 0$ for $a \to a_{max}$, where the diffusion ceofficient is locally maximal. The relaxation time, identified with the inverse of $\lambda_1$, tends to infinity, and evidence of \textit{critical slowing down} is found near these critical points. The limitations of this approximation arise from the form of the separation (\ref{sep}) of the exact function $x(t)$. While the direct separation of ``slow'' and ``fast'' motions is a convenient and simple tool for obtaining the reduced equation (\ref{VM}), its range of validity is limited.

We can  ask  whether the diffusion coefficient $D$ can be expressed in terms of the Lyapunov exponents $\lambda_1$ or $\lambda_2$. To this end, we propose the following test: from Eq. (\ref{root}) it follows that  
\begin{equation}
	\label{root2}
	|J_0(\psi)|  =  \lambda_1 \lambda_2 = -\lambda_1 (\lambda_1 + \gamma).    	
 \end{equation}
Next, we apply  the expression (\ref{strong}) for the diffusion coefficient $D$ to obtain the relation:   
\begin{equation}
	\label{strongD}
	D = \frac{Q}{\gamma I_0^2(\lambda_1 \lambda_2/Q)} = \frac{Q}{\gamma I_0^2(-\lambda_1 (\lambda_1 + \gamma)/Q)}, 
\end{equation}
where $I_0(x)$ is the modified Bessel function of the first kind \cite{specialF}.    
It is a suggested formula for the diffusion coefficient expressed in terms of the maximal Lyapunov exponent $\lambda_1$. We  have to 
verify whether Eq. (\ref{strongD}) reproduces the exact results of numerical analysis. In Fig. 1, panel (a), we compare it with the diffusion coefficient obtained from simulations of the Langevin equation (\ref{La}). The overall agreement is highly satisfactory. As in the case of the approximate Lyapunov exponent, exceptions occur for driving amplitudes $a$ close to their critical values $\{a_c, a_1, a_2, a_3\}$.  

Now, we can  explain in details how the approximate (red) curves  in Fig. 1 are obtained. The approximate diffusion coefficient 
is determined by Eq. (\ref{strongD}) in which the Lyapunov coefficient $\lambda_1$ is exact in the sense that it is  calculated  from numerical 
analysis of Eq. (\ref{deter}). The approximate maximal  Lyapunov exponent $\lambda_1 =\lambda_1(a)$ is obtained from Eq. (\ref{root}) with the argument  (\ref{psi}) of the function  $J_0(\psi) = J_0(\psi(a))$.

\section{Discussion} 

In the previous sections, we analyzed the similarity between the shapes of the diffusion coefficient $D=D(a)$ and the maximal Lyapunov exponent $\lambda_1= \lambda_1(a)$ as functions of the driving amplitude $a$. We also related $\lambda_1$ to the relaxation time and noted that as $\lambda_1$ approaches zero, the characteristic time $|1/\lambda_1|$ becomes increasingly long. Consequently, the spread of trajectories persists for a longer time, and the diffusion coefficient increases. Conversely, if $\lambda_1$ approaches a local minimum, $|1/\lambda_1|$ decreases, and the relaxation time of trajectories to periodic orbits is very short. Therefore, neighboring trajectories rapidly approach each other, which leads to a dwindling spread. Consequently, $D$ decreases. Of course, we must take into account the activation processes induced by thermal fluctuations and the jumps of particles to other trajectories as crucial components of the final value of $D$. Nevertheless, this simplified picture allows us to explain the main results presented in this paper. 

Thus far, we have fixed the parameters at a single point in the three-dimensional space $\{\gamma, \omega, Q\}$. Is this an exceptional set of values for which the correlation between $D$ and $\lambda_1$ emerges? To answer this question, we explored a selected region of the parameter space to locate such a correlation and found that it is quite robust with respect to small parameter changes, particularly when the damping constant $\gamma$ is varied. In Fig. 4, we depict the influence of $\gamma$. One can observe that when $\gamma$ is slightly varied, the correlation between $D$ and $\lambda_1$ is preserved, but the shape of the function $\lambda_1(a)$ changes more drastically: the intervals of the driving amplitude $a$ in which $\lambda_1$ is minimal become larger, and within these intervals, the diffusion coefficient is small. As a general rule, the following applies: if $\lambda_1$ decreases (i.e., becomes more negative), $D$ also decreases; if $\lambda_1$ increases (i.e., approaches zero), $D$ also increases. Finally, if $\gamma$ decreases to values at which running trajectories emerge, this correlation is destroyed, and the Lyapunov exponent no longer shares the features exhibited by the diffusion coefficient.

\begin{figure}[b]
\centering
\includegraphics[width=0.8\linewidth]{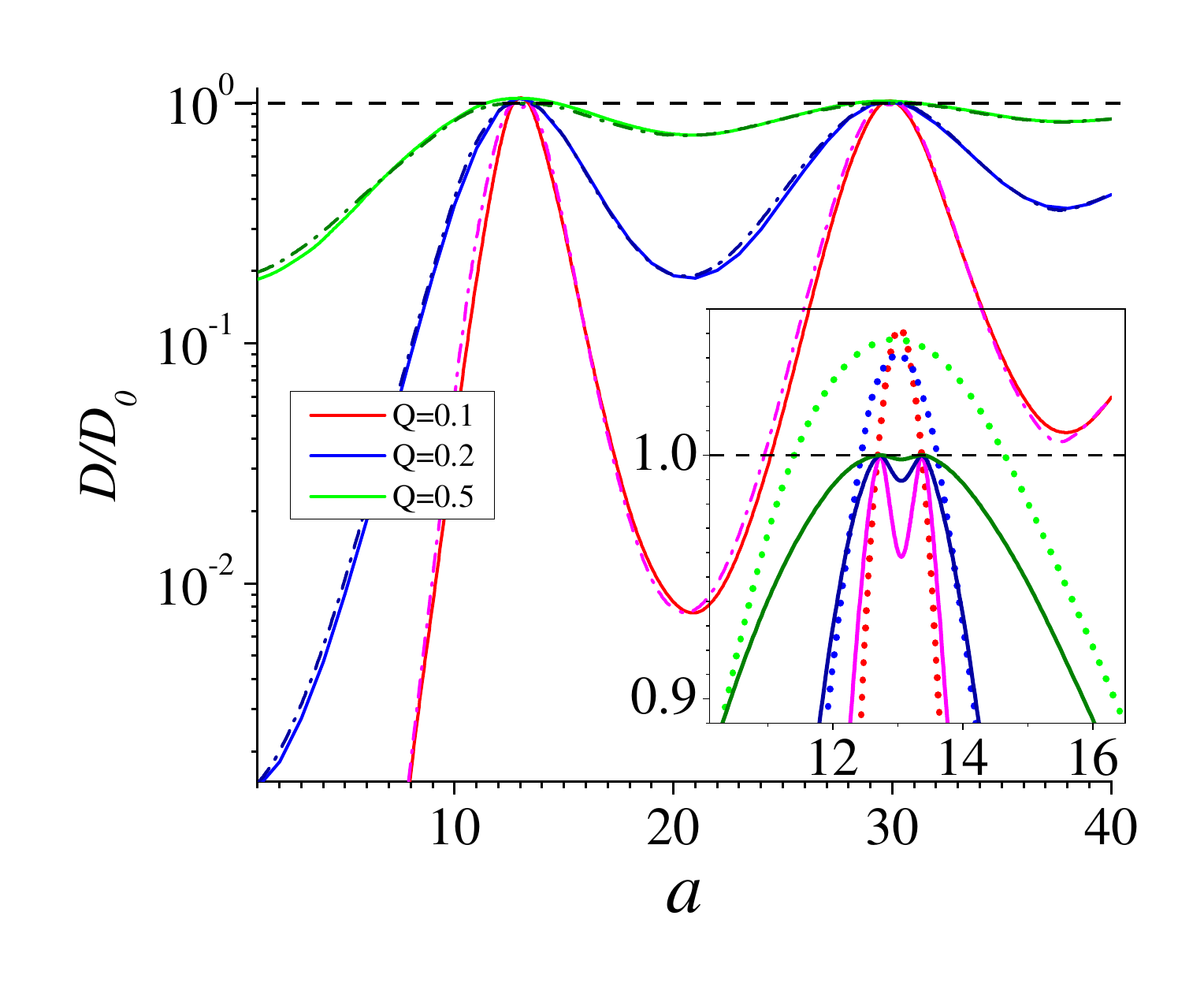}
\caption{Influence of temperature $Q$ on the diffusion coefficient. The doted  curves  are obtained from  simulations of the Langevin equation (\ref{La}), whereas  continues -  from Eq. (\ref{strongD}). Other parameters are the same as in Fig. \ref{fig1}.}
\label{fig5}
\end{figure}

For completeness of results, we show the influence of temperature on  the diffusion coefficient $D$. In Fig. 5, three values of  dimensionless temperature $Q$ are 
assumed: $Q=0.2$ is  twice as higher as $Q=0.1$ assumeed in the main part of the paper and $Q=0.5$ is five times higher than $Q=0.1$.  Of course, it does not influence behavior of the Lyapunov exponents of the deterministic system  given by 
 Eq. (\ref{deter}), however, the diffusion coefficient  of the noisy system (\ref{La}) depends on $Q$. The first observation is that  $D$ is greater if temperature is higher. It seems to be obvious but in a general case it is not the rule \cite{myPRE}.  The second observation is that the region of the driving amplitude $a$  in which  $D>D_0$ is wider for higher temperature. 
 We can conclude that the correlation between the diffusion coefficient and the maximal Lyapunov exponent is insensitive to small changes of temperature and still is preserved even if temperature is five times higher than the initial one. The theoretical results for the diffusion coefficient $D$ based on Eq. (\ref{strongD}) reproduce the simulation results but  a discrepancy is observed in the intervals of $a$  in which the center of  periodic orbits of the deterministic system (\ref{deter}) 
 is relocated from the minimum of the periodic potential $U(x)$ to its  maximum.  This drawback could   be 
 improved if the expression for the diffusion coefficient having a wider  range of validity  than Eq. (\ref{strong}) is applied.

\section*{Appendix A: Scaling of equations}

Here, we present the dimensional form of the Langevin equation (\ref{La}). We consider a classical Brownian particle of mass $M$ moving in a one-dimensional, spatially periodic potential $W(x)$, subjected to an unbiased and symmetric time-periodic force $F(t)$. The corresponding Langevin equation is given by \cite{spiechowicz2016njp}:
\begin{equation}
	\label{model}
	M\ddot{x} + \Gamma\dot{x} = -W'(x) + F(t) + \sqrt{2\Gamma k_B T}\,\xi(t),
\end{equation}
where the dot and the prime denote differentiation with respect to time $t$ and the particle coordinate $x$, respectively. The parameter $\Gamma$ represents the friction (damping) coefficient. The symmetric potential $W(x) = W(x+L)$ with spatial period $L$ and barrier height $2 \Delta W$ is given by: 
\begin{equation}
	\label{potential}
	W(x) =  -\Delta W\cos{\left( \frac{2\pi}{L}x \right)}.
\end{equation}
The external ac force with amplitude $F_0$ and angular frequency $\Omega$ is assumed to take the simplest harmonic form, namely:   
\begin{equation}
	F(t) = F_0 \sin{(\Omega t +\phi_0)}. 
\end{equation}
Thermal equilibrium fluctuations arising from the interaction of the particle with a heat bath at temperature $T$ are modeled as $\delta$-correlated Gaussian white noise with zero mean, 
\begin{equation}
	\langle \xi(t) \rangle = 0, \quad \langle \xi(t)\xi(s) \rangle = \delta(t - s), 
\end{equation}
where the angular brackets $\langle \cdot \rangle$ denote an ensemble average over the white noise realizations. The noise intensity $2\Gamma k_B T$ in Eq. (\ref{model}) is dictated by the fluctuation-dissipation theorem \cite{kubo1966}, with $k_B$ being the Boltzmann constant. If $A=0$, the stationary state of the system is a thermal equilibrium state. Conversely, if $A \neq 0$, the external force $F(t)$ drives the system out of equilibrium. 

Next, we transform Eq. (\ref{model}) into its dimensionless form. To this end, we define the dimensionless length $\hat x$ and time $\hat t$ as: 
\begin{equation}
	\label{scaling}
	\hat{x} = 2\pi \frac{x}{L}, \quad \hat{t} = \frac{t}{\tau_0}, \quad \tau_0 = \frac{L}{2\pi}\sqrt{\frac{M}{\Delta W}}.
\end{equation}
In terms of these new variables, Eq. (\ref{model}) takes the form:
\begin{equation}
	\label{dimless-model}
	\ddot{\hat{x}} + \gamma\dot{\hat{x}} = -\sin{\hat{x}} + a \sin (\omega \hat{t}) +  \sqrt{2\gamma Q} \hat{\xi}(\hat{t}).
\end{equation}
Under this scaling, the dimensionless mass is $m = 1$, and the remaining four dimensionless parameters are defined as follows:
\begin{eqnarray}
	\gamma = \frac{\tau_0}{\tau_1}, \quad a = \frac{1}{2\pi}\frac{L}{\Delta W} F_0,  
\quad \omega = \tau_0 \Omega, \quad Q = \frac{k_B T}{\Delta W}, \nonumber \\
\end{eqnarray}
where the second characteristic time is $\tau_1 = M/\Gamma$. This parameter represents the velocity relaxation time of a free Brownian particle. In contrast, the characteristic time $\tau_0$ corresponds to the period of small oscillations at the bottom of the potential wells. 

The rescaled potential with period $L=2\pi$ is $\hat{U}(\hat{x}) = W((L/2\pi)\hat{x})/\Delta W = -\cos{\hat x}$, and the corresponding potential force is $-\hat{U}'(\hat{x})=-\sin \hat{x}$. The rescaled thermal noise is given by $\hat{\xi}(\hat{t}) = (L/2\pi \Delta W)\xi(t) = (L/2\pi \Delta W)\xi(\tau_0\hat{t})$ and shares the same statistical properties as $\xi(t)$, meaning $\langle \hat{\xi}(\hat{t}) \rangle = 0$ and $\langle \hat{\xi}(\hat{t})\hat{\xi}(\hat{s}) \rangle = \delta(\hat{t} - \hat{s})$. The dimensionless noise intensity $Q$ is the ratio of the thermal energy to half the activation energy required for the particle to overcome the unscaled potential barrier. Hereafter, we will use these dimensionless variables exclusively. To simplify the notation further, we omit the hat symbols in Eq. (\ref{dimless-model}).

\section*{Appendix B: Details of the simulations}

The analysis of the Langevin equation (\ref{La}) was based on extensive numerical simulations using the Compute Unified Device Architecture (CUDA) environment on a modern desktop Graphics Processing Unit (GPU). This approach provided an acceleration of approximately three orders of magnitude compared to standard Central Processing Unit (CPU) implementations \cite{spiechowicz2015cpc}. The Langevin equation (\ref{La}) was solved numerically utilizing a second-order predictor-corrector algorithm \cite{platen} with an integration time step of $h = 10^{-2} \times 2\pi/\omega$, where $2\pi/\omega$ denotes the period of the external driving force $a\cos{(\omega t)}$. The quantities describing the diffusive properties of the system were determined by averaging over an ensemble of $N = 2^{14} = 16384$ trajectories. Each trajectory was assigned independent initial conditions $x(0)$ and $v(0)$, sampled uniformly from the intervals $[0, 2\pi]$ and $[-2, 2]$, respectively. The simulation time typically spanned up to $10^6$ driving periods, which proved sufficient to reach the asymptotic long-time regime, as confirmed by the stationarity of the diffusion coefficient.  

 For Eq. (\ref{deter}), we computed Lyapunov exponents using the Benettin algorithm, integrating both the differential  equations and variational equations. 
 The  QR decomposition (Gram-Schmidt orthonormalization) was applied at adaptive intervals to maintain vector independence. This enabled high-precision estimation of Lyapunov spectra in continuous dissipative systems, following the method described in Ref. [\onlinecite{benet}].

\section*{Appendix C: Vibrational mechanics approach}

There are several approximate methods for the analysis of systems driven by time-periodic forces, such as the method of averaging, the method of multiple scales, vibrational mechanics, and the method of direct separation of motions. Here, we follow the vibrational mechanics approach \cite{sorokin}, which is applicable at high driving frequencies. For the dynamics governed by Eq. (\ref{La}) or Eq. (\ref{deter}), the trajectory $x(t)$ is represented in the form \cite{boroEPL,march-prl}: 
\begin{equation}
	\label{sep}
	x(t) = {\tilde x}(t) + \psi \sin(\omega t +\phi_1), 
\end{equation}
where  $\tilde x(t)$ represents a slowly time-modulated process. The fast oscillating terms can be averaged over the period of the time-periodic force  and  then  the Langevin equation (\ref{La})  for the variable $\tilde x(t) \equiv \tilde x$ takes the form 
\begin{equation}
	\label{VM}
	\ddot{\tilde x} + \gamma\dot{\tilde x} = -J_0(\psi) \sin{\tilde x} + \sqrt{2\gamma Q} \, \xi(t), 
\end{equation}
where  $J_0(\psi)$ is the zero order Bessel function \cite{specialF} of the argument  
\begin{equation}
	\label{psi}
	\psi = \psi(a) = \frac{a}{\omega \sqrt{\omega^2 + \gamma^2}}. 
\end{equation}
We now want to apply Eq. (\ref{VM}) to obtain both the diffusion coefficient and for $Q=0$ the Lyapunov exponent. We note that the starting total time-dependent potential ${V}(x, t) = -\cos(x) -ax \sin(\omega t+\phi_0)$ is replaced by the time-independent renormalized potential ${V}_{\mbox{eff}}(x) = -J_0(\psi) \cos(x)$. In this approach, the particle diffuses in a renormalized potential of the effective barrier $2J_0(\psi)$. 
For strong damping, the approximate expression for the diffusion coefficient   takes  the following form \cite{lifson} 
\begin{equation}
	\label{strong}
	D = \frac{Q}{\gamma I_0^2(J_0(\psi)/Q)}, 
\end{equation}
where $I_0(x)$ is the modified Bessel function of the first kind \cite{specialF}.

\section*{Acknowledgement}
This work was supported in part by the National Science Centre (NCN) Grant No. 2024/54/E/ST3/00257 (J.S.) and  PLGrid Infrastructure under grant no. PLG/2025/018093 (I.G.M.).   \mbox{I. G. M.} acknowledges University of Silesia for the hospitality since the beginning of the war,  February 24, 2022.  \\

\noindent {\bf Author contributions} \\ 
I.G.M. coordinated  the  study;  conceptualization, discussion and editing of the manuscript: I.G.M., J.S. and J.{\L.};   I.I.M and D.I.  contributed to the design of the numerical solution schemes and to the development of numerical codes, and performed all numerical calculations and 
simulations. All authors contributed to writing the paper and gave final approval for publication. \\

\noindent {\bf Competing interests } \\
The authors declare no competing interest  \\

\noindent {\bf Additional information } \\
Correspondence and requests for materials should be addressed to J.Ł. (e-mail: jerzy.luczka@us.edu.pl)

\end{document}